\documentclass[aip,preprint]{revtex4-1}

\usepackage{graphicx}
\usepackage{multirow}
\usepackage{amsmath}
\usepackage{lineno}
\usepackage{figcaps}
\begin{document}

\title{Gafchromic film dosimetry: calibration methodology and error analysis} 
\author{W. Crijns}\email{wouter.crijns@uzleuven.be}\author{F. Maes}\author{F. Van den Heuvel}
\affiliation{Department of Radiotherapy, University Hospital Leuven, Belgium}
\date{\today}
\begin{abstract}
\textbf{Purpose: }To relate the physical transmittance parameters of the water equivalent Gafchromic EBT 2 film with the delivered dose in a transparent absolute calibration protocol. The protocol should be easy to understand, easy to perform, and should be able to predict the residual dose error.

\textbf{Methods: }
The protocol uses at most two uncut calibration films for a whole batch. The calibration films are irradiated on different positions with static fields with varying dose levels.

The transmittance trough the film, $T$, is preferred over the optical density. An analytical function is used to associate $T$ with the delivered dose based on physical characteristics of the film: the minimal and maximal $T$-values, $T_0$ and $T_\infty$, and a factor scaling the dose, $\beta_3$.

The dose uncertainty of the protocol is calculated using an error propagation analysis of the calibration curve. This analysis makes use of the small scale and the large scale variations in the film response (local and global uniformity).
Before calibrating the film, the stability of unirradiated films is evaluated. The relevance of the positions of the different dose levels is studied.

\textbf{Results: }
Both a production and a spatial dependency of the unirradiated films was noticed. This spatial dependency required an avoidance of a quarter of the film. The transmittance is also affected by the storage and the transport conditions.

The lowest residual errors and the highest significance for $T_0$, $T_\infty$, and $\beta_3$ was achieved with arbitrary dose levels for different positions on the film. Two calibration films give the better results. Significant calibration curves were found for both the red and the green color channel.

Large differences ($>0.02$Gy) between calibration curves, with different positions for the dose levels are seen.

In the [0.04,2.5]Gy dose range the red channel calibration curve has dose errors ranging from -2.3\% to 4.9\%. For the green channel the error range is [-3.6\%,6.9\%].

The calculated error propagation was able to predict an upper limit for the red channel dose errors. For the green channel this prediction was not achieved.

\textbf{Conclusions: }The gafchromic EBT2 films are properly calibrated with an accessible robust calibration protocol. The protocol largely deals with the uniformity problems of the film. The proposed method allowed to relate the dose with the red channel transmittance using only $T_0$, $T_\infty$, and a dose scaling factor. Based on the local and global uniformity the red channel dose errors could be predicted to be smaller than 5\%.
\end{abstract}
\maketitle
\section{INTRODUCTION}
Radiochromic films, such as Gafchromic EBT and EBT 2 films\cite{ISP}, are a popular tool to measure dose depositions from ionizing radiation. 
Most of the previously published studies\cite{Rink_05,Rink_05b,Todorovic06,Rink07b,Todorovic06,Andres_10} have focused on the gafchromic EBT film, the predecessor of the EBT 2 film. However, the active layer of both types of films is the same, which results in comparable outcomes for the EBT and the EBT 2 film \cite{Butson_09,Andres_10}.
The main advantages of these type of films are the fact that these films have the following properties:
\paragraph{}A near flat response to different photon energies \citep{Todorovic06,Rink07b}
\paragraph{}Close to water equivalence (Z$_{\text{eff EBT 2}}$=6.98 \citep{Todorovic06})
\paragraph{}A thickness less then 0.3mm\cite{Rink_05,Rink_05b}
\paragraph{}A low perturbation of the irradiated media, which is certainly important for the evaluation of recent rotating treatment techniques like VMAT\cite{Otto_08}
\paragraph{}A reasonably large dynamic range ([0,10]Gy\cite{Andres_10}, [0.1,8]Gy\cite{Todorovic06})
\paragraph{}A high resolution (0.22$\mu$m\cite{Rink_05,Rink_05b})

Disadvantages of the films are their non-uniformity and production stability, which has an impact on the choice between transmittance ($T$), and optical density ($OD$), as dose dependent quantity (section \ref{S_ODorT}).

However, the combination of the afore mentioned properties makes gafchromic EBT 2 film, further referred to as film, usable in almost every application in radiotherapy.
For example, at the University Hospital Leuven the films are clinically used : for in-vivo measurements of total body and electron skin irradiations; for 
Volumetric Modulated Arc Therapy quality assurance; for internal audits of brachytherapy, and stereotactic treatments; for the introduction of new treatment techniques\citep{Crijns_AAPM09}, and new clinical trials (FLAME trial\cite{FLAME_trial}).

Additionally, there is a large range of research purposes such as; out of field dosimetry \cite{Defraene_Estro11}, estimation of microscopic dose distributions in nano-particle enhanced radiation therapy\citep{Heuvel_Astro10,Heuvel_Estro11}, and the optimization of the dosimetry of fluoroscopic CT protocols.

The wide use of these film requires a transparent and accessible absolute calibration protocol. The calibration methodology aims to use one or two uncut calibration films to represent a whole batch. The use of an uncut film avoids puzzling with calibration film fragments, and avoids sharp transmission and diffraction edges in the film scanner transmission system. Calibration films and patient quality assurance (QA) films are always scanned successively, but they are not necessarily irradiated on the same day. An error propagation analysis of the calibration protocol is built using the small scale and the large scale variations in the film response. 

\section{MATERIALS AND METHODS}
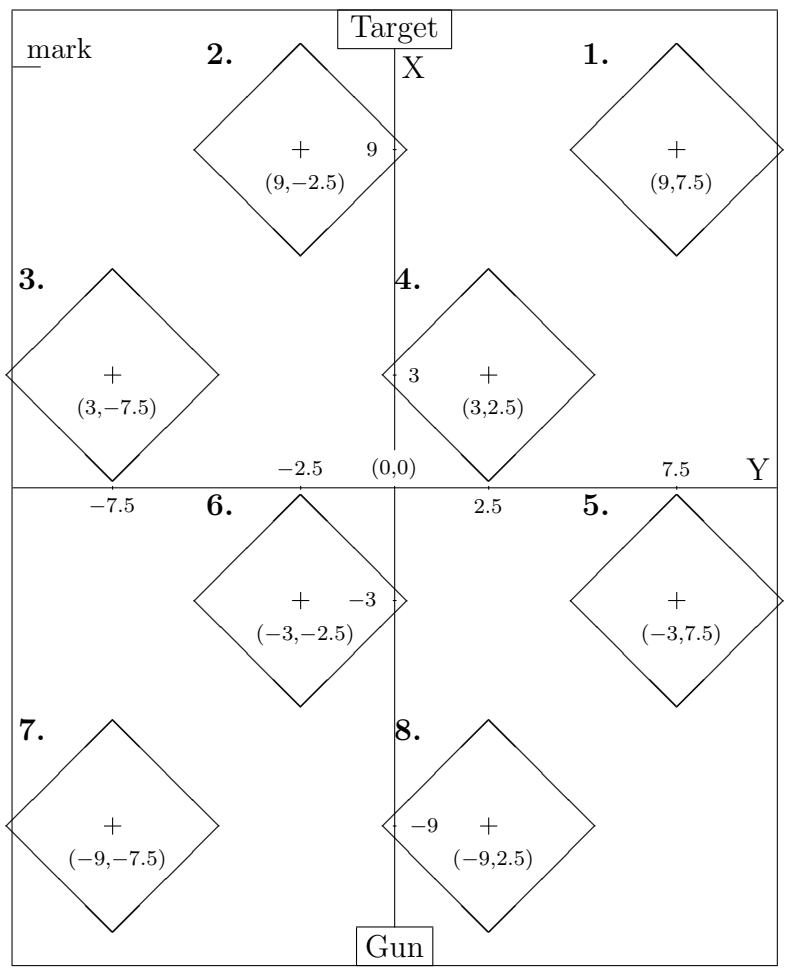
\begin{figure}
    \begin{center}
\setlength{\unitlength}{0.5mm}
\begin{picture}(203.2,254)
\put(0,0){\framebox(203.2,254)} \put(0,127){\line(1,0){203.2}}
\put(91.6,0){\framebox(20,10)[c]{Gun}}\put(101.6,10){\line(0,1){117.5}}
\put(101.6,137){\line(0,1){107}}
\put(86.6,244){\framebox(30,10)[c]{Target}}
\put(101.6,234){\makebox(10,10)[c]{X}}
\put(193.2,127){\makebox(10,10)[c]{Y}}
\put(0,239){\line(1,0){7.5}}\put(7.5,239){\makebox(10,10)[c]{\small{mark}}}
\put(96.6,127){\makebox(10,10)[c]{$_{(0,0)}$}}
\put(26.6,126.5){\line(0,1){1}}
\put(21.6,117){\makebox(10,10)[c]{$_{-7.5}$}}
\put(76.6,126.5){\line(0,1){1}}
\put(71.6,127){\makebox(10,10)[c]{$_{-2.5}$}}
\put(126.6,126.5){\line(0,1){1}}
\put(121.6,117){\makebox(10,10)[c]{$_{2.5}$}}
\put(176.6,126.5){\line(0,1){1}}
\put(171.6,127){\makebox(10,10)[c]{$_{7.5}$}} 
\put(101.1,217){\line(1,0){1}}
\put(94.1,212){\makebox(10,10)[l]{$_{9}$}}
\put(101.1,157){\line(1,0){1}}
\put(98.6,152){\makebox(10,10)[r]{$_{3}$}}
\put(101.1,97){\line(1,0){1}}
\put(89.1,92){\makebox(10,10)[l]{$_{-3}$}}
\put(101.1,37){\line(1,0){1}}
\put(103.6,32){\makebox(10,10)[r]{$_{-9}$}}
\put(148.32,188.72){\begin{picture}(56.56,56.56)
\put(0,0){\makebox(56.56,56.56)[lt]{\textbf{ 1.}}}
    \put(0,28.28){\line(1,1){28.28}}
    \put(0,28.28){\line(1,-1){28.28}}
    \put(56.56,28.28){\line(-1,1){28.28}}
    \put(56.56,28.28){\line(-1,-1){28.28}}
    \put(14.14,16.5){\makebox(28.28,28.28)[cb]{ $_{(9,7.5)}$}}
    \put(28.28,26){\line(0,1){4.56}}
    \put(26,28.28){\line(1,0){4.56}}
    \end{picture}}
\put(48.32,188.72){\begin{picture}(56.56,56.56)
\put(0,0){\makebox(56.56,56.56)[lt]{\textbf{ 2.}}}
    \put(0,28.28){\line(1,1){28.28}}
    \put(0,28.28){\line(1,-1){28.28}}
    \put(56.56,28.28){\line(-1,1){28.28}}
    \put(56.56,28.28){\line(-1,-1){28.28}}
    \put(14.14,16.5){\makebox(28.28,28.28)[cb]{ $_{(9,-2.5)}$}}
    \put(28.28,26){\line(0,1){4.56}}
    \put(26,28.28){\line(1,0){4.56}}
    \end{picture}}
\put(-1.68,128.72){\begin{picture}(56.56,56.56)
\put(0,0){\makebox(56.56,56.56)[lt]{\textbf{ 3.}}}
    \put(0,28.28){\line(1,1){28.28}}
    \put(0,28.28){\line(1,-1){28.28}}
    \put(56.56,28.28){\line(-1,1){28.28}}
    \put(56.56,28.28){\line(-1,-1){28.28}}
    \put(14.14,16.5){\makebox(28.28,28.28)[cb]{ $_{(3,-7.5)}$}}
    \put(28.28,26){\line(0,1){4.56}}
    \put(26,28.28){\line(1,0){4.56}}
    \end{picture}}
\put(98.32,128.72){\begin{picture}(56.56,56.56)
\put(0,0){\makebox(56.56,56.56)[lt]{\textbf{ 4.}}}
    \put(0,28.28){\line(1,1){28.28}}
    \put(0,28.28){\line(1,-1){28.28}}
    \put(56.56,28.28){\line(-1,1){28.28}}
    \put(56.56,28.28){\line(-1,-1){28.28}}
    \put(14.14,16.5){\makebox(28.28,28.28)[cb]{ $_{(3,2.5)}$}}
    \put(28.28,26){\line(0,1){4.56}}
    \put(26,28.28){\line(1,0){4.56}}
    \end{picture}}
\put(148.32,68.72){\begin{picture}(56.56,56.56)
\put(0,0){\makebox(56.56,56.56)[lt]{\textbf{ 5.}}}
    \put(0,28.28){\line(1,1){28.28}}
    \put(0,28.28){\line(1,-1){28.28}}
    \put(56.56,28.28){\line(-1,1){28.28}}
    \put(56.56,28.28){\line(-1,-1){28.28}}
    \put(14.14,16.5){\makebox(28.28,28.28)[cb]{ $_{(-3,7.5)}$}}
    \put(28.28,26){\line(0,1){4.56}}
    \put(26,28.28){\line(1,0){4.56}}
    \end{picture}}
\put(48.32,68.72){\begin{picture}(56.56,56.56)
\put(0,0){\makebox(56.56,56.56)[lt]{\textbf{ 6.}}}
    \put(0,28.28){\line(1,1){28.28}}
    \put(0,28.28){\line(1,-1){28.28}}
    \put(56.56,28.28){\line(-1,1){28.28}}
    \put(56.56,28.28){\line(-1,-1){28.28}}
    \put(14.14,16.5){\makebox(28.28,28.28)[cb]{ $_{(-3,-2.5)}$}}
    \put(28.28,26){\line(0,1){4.56}}
    \put(26,28.28){\line(1,0){4.56}}
    \end{picture}}
\put(-1.68,8.72){\begin{picture}(56.56,56.56)
\put(0,0){\makebox(56.56,56.56)[lt]{\textbf{ 7.}}}
    \put(0,28.28){\line(1,1){28.28}}
    \put(0,28.28){\line(1,-1){28.28}}
    \put(56.56,28.28){\line(-1,1){28.28}}
    \put(56.56,28.28){\line(-1,-1){28.28}}
    \put(14.14,16.5){\makebox(28.28,28.28)[cb]{ $_{(-9,-7.5)}$}}
    \put(28.28,26){\line(0,1){4.56}}
    \put(26,28.28){\line(1,0){4.56}}
    \end{picture}}
\put(98.32,8.72){\begin{picture}(56.56,56.56)
\put(0,0){\makebox(56.56,56.56)[lt]{\textbf{ 8.}}}
    \put(0,28.28){\line(1,1){28.28}}
    \put(0,28.28){\line(1,-1){28.28}}
    \put(56.56,28.28){\line(-1,1){28.28}}
    \put(56.56,28.28){\line(-1,-1){28.28}}
    \put(14.14,16.5){\makebox(28.28,28.28)[cb]{ $_{(-9,2.5)}$}}
    \put(28.28,26){\line(0,1){4.56}}
    \put(26,28.28){\line(1,0){4.56}}
    \end{picture}}
\end{picture}
\caption{Schematic overview of the calibration image. The outer
rectangle represents the film. All length indication are in
[cm].\label{F_Calscheme}}
    \end{center}
\end{figure}

The calibration protocol is based on eight static fields irradiated on an uncut 8$\times$10 inch film, see figure \ref{F_Calscheme}. Table \ref{T_FilmSumm} lists all the films used in this work.

\begin{table*}\caption{Summary of the different irradiated and blank films used in this work.\label{T_FilmSumm}}
\begin{center}
\begin{small}
\begin{tabular}{c|c|c|l}
  \hline \hline
 Experiment  & Film & Batch Nr. & Description \\ \hline\hline
 \multirow{2}{*}{Storage} & $V_{1-7}$  & A081610 & Vacuum packed films without interleaving tissue. \\
  &$V_{7-15}$ & A081610 & Vacuum packed films with interleaving tissue.\\ \hline
 \multirow{15}{*}{Calibration} & $E_{1}$    & F03161001 & Calibration film with: $MU=[33, 66, 100, 125, 150, 175, 200, 225]$ and $D_{i,i\text{Cal 100 MU}}=1.0644Gy$. \\
  &$E_{2,3}$  & F03161001 & Calibration film with: $MU=[33, 66, 100, 125, 150, 175, 200, 225]$ and $D_{i,i\text{Cal 100 MU}}=1.0689Gy$. \\
  &$E_{4}$    & F03161001 & Calibration film with the opposite MU order:\\&&& $MU=[225, 200, 175, 150, 125, 100, 66, 33]$ and$D_{i,i\text{Cal 100 MU}}=1.0689Gy$. \\
  &$E_{5}$    & F03161001 & Calibration film with a random MU order and $D_{i,i\text{Cal 100 MU}}=1.0689Gy$; \\
  &                       && segment 1 to 8: $MU=[150, 33, 225, 66, 200, 100, 125, 175]$). \\
  &$E_{6}$    & F03161001 & The film is irradiated with $\pm$ equal number of MU's per row \\
  &                        && $D_{i,i\text{Cal 100 MU}}=1.0689Gy$;\\
  &                        && segment 1 to 8: $MU=[33, 225, 200, 66, 100, 175, 150, 125]$). \\
  &$E_{7}$    & F03161001 & The film is uniformly irradiated with a 30$\times$30 field, 2.38Gy at the central axis\\
  &$E_{8,9,11}$&F03161001 & These films are left blank. \\
  &$E_{10}$    & F03161001 & Calibration film with the opposite MU order of $E_{1}$ and \\
  &                       && $D_{i,i\text{Cal 100 MU}}=1.0689Gy$ \\
  &                       && this is the same order as $E_4$, but segment five was left blank,\\
  &                       && and segment six got $100+125$MU;\\
  &                       && segment 1 to 8: $MU=[225, 200, 175, 150, 0, 225, 66, 33]$)\\\hline
 Stability &$B_{1-13}$ & variate   & Blank films used in clinical practice through out 2010, see figure \ref{F_stability} and \ref{F_stabilityU}. \\
  \hline \hline
\end{tabular}
\end{small}
\end{center}
\end{table*}
To relate $T$ with the dose, $D$, only three parameters are used ($\beta_1,\beta_2,\beta_3$). These parameters can be reduced to the zero dose transmittance, $T_0$, the infinite dose transmittance, $T_{\infty}$, and a factor scaling the dose, $\frac{1}{\beta_3}$.

\subsection{Theory\label{S_theory}}
\subsubsection{Optical density versus transmittance \label{S_ODorT}}
Historically, OD is the quantity of choice because of its linear relation with the dose delivered, e.g. TG 55 reports a linear net OD response from 0 to 30 Gy, and the from 30 to 100Gy for the MD-55-2 radiochromic film\cite{TG55_1998}. Such a linear relation allows a scaling of relative dose measurements. For the gafchromic EBT film, \citeauthor{Rink_05} mentioned a nonlinear OD response\cite{Rink_05}.

An exponential function can be used to describe such a nonlinear relation between the dose and OD\citep{vanBattum06,Zhu_03,Jeffrey_81} , e.g. OD$=A\cdot\left(1-e^{-B\cdot D}\right)$. Such an expression results in zero optical density, when no dose is applied, OD$_{0\text{Gy}}=0$.
On the other hand, the optical density is defined as OD$ = \log_{10}\left(\frac{\phi_i}{\phi_t}\right)$ (\cite{Anderson84}, formula 12.34). Where $\phi_i$ is the visible light fluence incident on the film and $\phi_t$ is the fluence transmitted through the film. To reconcile both expressions it is necessary to replace $\phi_i$ by the fluence transmitted through an unirradiated film, $\phi^*_i$, which results in zero OD when no dose is delivered (OD$_{0\text{Gy}}=0$).

\begin{figure}
 \includegraphics[width=7cm]{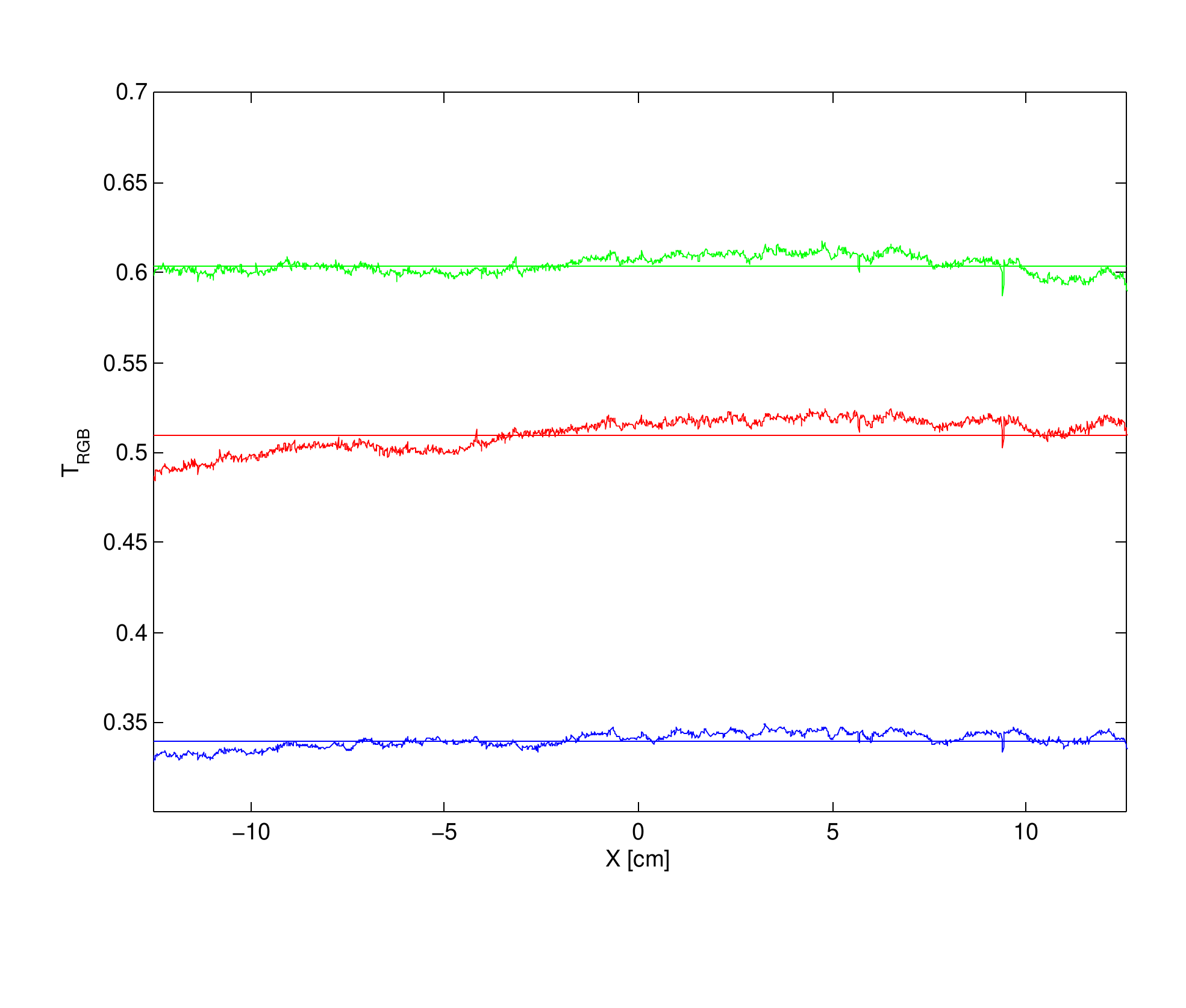}%
 \caption{The transmittance ($T_{\text{RGB}}^*$) allonge the central pixel line, of a uniformly irradiated film (table \ref{T_FilmSumm}, Film$_{E_{7}}$). A reduction of the measured signal is noticed for $X<-5$cm.\label{F_linescan}}%
\end{figure}

Over the course of 2010 we had variable success using the scan of a separate unirradiated film as an estimate of $\phi^*_i$ (table \ref{T_FilmSumm} Film$_{B_1-B_{13}}$). Using this approach two main problems, were noticed. Firstly, a stability problem; the unirradiated film, and therefore the estimated $\phi^*_i$-value, changes over (production)--time. And secondly $\phi^*_i$ is spatial-dependent, see figure \ref{F_linescan}. This non-uniformity requires a location dependent $\phi^*_i$-estimation.
Because a stable estimation method for $\phi^*_i$ was not obvious, and because the loss of linearity for the OD-Dose response, our research concentrates on a straightforward association between the physical quantities transmittance ($T=\frac{\phi_t}{\phi_i}$) and dose.

\subsubsection{Association between transmittance and delivered dose}
\begin{figure}
 \includegraphics[width=7cm]{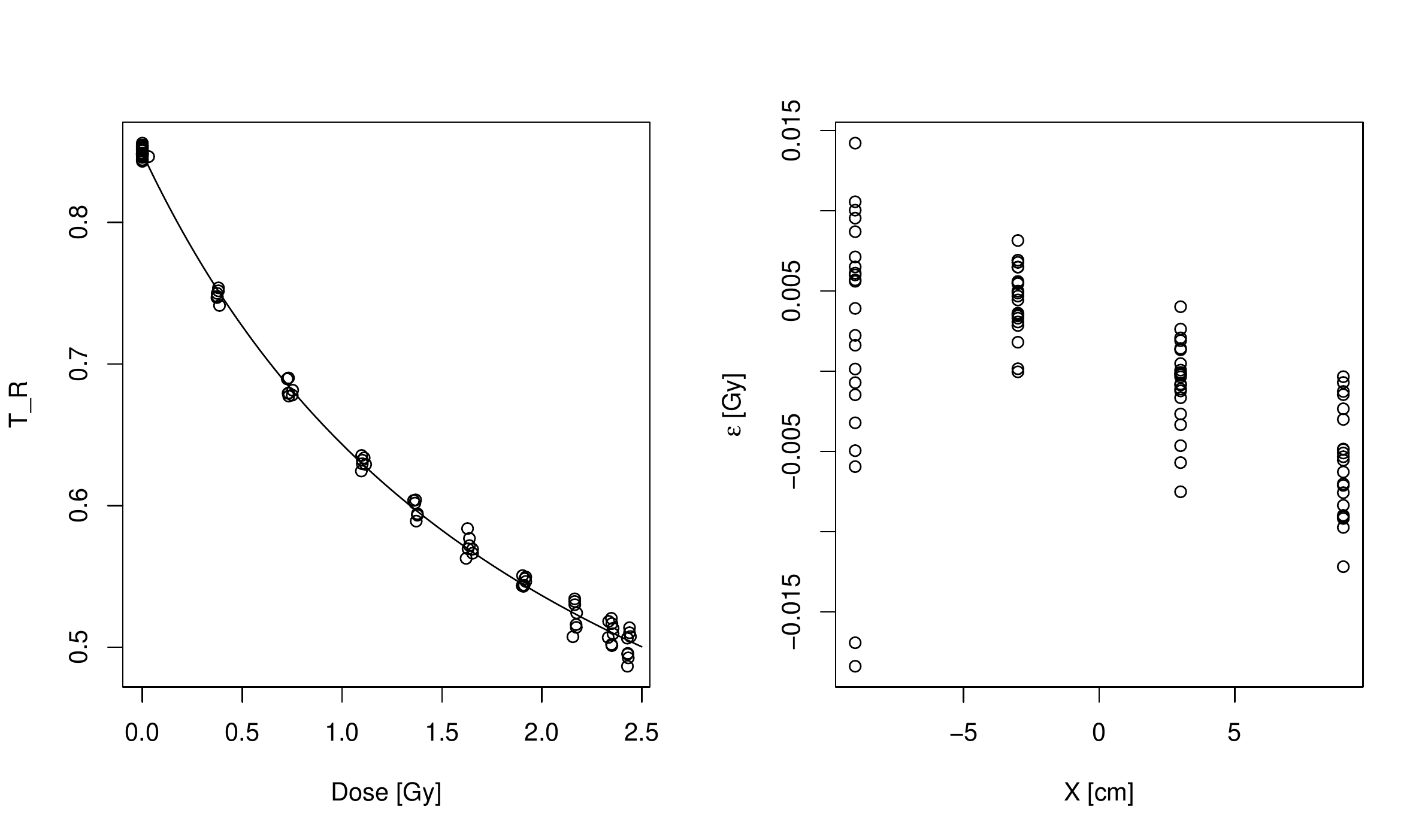}%
 \caption{A fit of the red color channel data of all the films: Film$_{E_1-E_{11}}$. On the left the data is shown together with the fit $T_R = \frac{1.833 + 0.199\cdot D_{\text{IC}}}{2.159+D_{\text{IC}}}$. On the right the residuals are shown in function of their X location on the film. Note the X-dependence of the residuals as well as the larger spread of the residuals of the segments with $X=-9cm$\label{F_AllFit}}%
\end{figure}

Figure \ref{F_AllFit} illustrates the association between the dose delivered and the transmittance as measured by us.
Mathematically, $T$ decreases monotonically as $D$ increases. Such an association can be described by a rational function\cite{Micke_11}, e.g. $T = \frac{\beta_1 + \beta_2 \cdot D}{\beta_3 + D}$.  This function is more interesting than a polynomial function due to the absence of points of inflection. The rational function is extended with a term depending on the pixel location, the $\beta_4 \cdot X$-term in equation \ref{E_ItoDose}.  This term takes into account a non-uniformity along the 10 inch side of the film. The transitions from equation \ref{E_ItoDose} to \ref{E_DosetoI} can be found in appendix \ref{S_A1}.
\begin{subequations}
\begin{equation}
T=\frac{ \beta_1 + \beta_2 \cdot D }{\beta_3 + D} + \beta_4 \cdot X \label{E_ItoDose}
\end{equation}
\begin{equation}
D=\frac{\left(\beta_3\cdot T-\beta_3\cdot\beta_4 \cdot X \right) - \beta_1}{\beta_2 - T+\beta_4 \cdot X }\label{E_DosetoI}
\end{equation}
\end{subequations}
Ignoring the non-uniformity term, $\beta_4=0$, equation \ref{E_ItoDose} can be interpreted as follows.
\paragraph*{$T_{0} = \frac{\beta_1}{\beta_3}$ :}The transmittance when no dose is delivered equals the ratio of $\beta_1$ and $\beta_3$.
\paragraph*{$T_{\infty}=\beta_2$ :}$\beta_2$ is the transmittance resulting from a theoretical infinite dose, $T_\infty=\lim_{D\rightarrow\infty}\left(\frac{\beta_1 + \beta_2 \cdot D}{\beta_3 + D}\right)$ (l'H\^opital's rule).
In appendix \ref{S_A1},equations \ref{E_ItoDose} and \ref{E_DosetoI} are rewritten to equations \ref{E_ItoDose2} and \ref{E_DosetoI2}, illustrating the relation of the physical quantities $T$, $T_0$, $T_\infty$, and $D$. From these equations, $\frac{1}{\beta_3}$ can be seen as a factor scaling the impact of the dose.
\begin{subequations}
\begin{equation}
T=T_\infty + \frac{T_0 - T_\infty }{1+\frac{D}{\beta_3}}  + \beta_4 \cdot X \label{E_ItoDose2}
\end{equation}
\begin{equation}
D=\beta_3 \cdot \frac{\left(T_0 + \beta_4 \cdot X\right)-T}{T-\left(T_\infty+\beta_4\cdot X\right)}\label{E_DosetoI2}
\end{equation}
\end{subequations}

\subsubsection{Error analysis}
The partial derivatives of equation \ref{E_DosetoI} are used in an error propagation analysis.
This analysis estimates the error made when film measurements are performed, using the proposed calibration methodology.
The dose uncertainty of the protocol, $\sigma_D$, is calculated using estimations of the position variance $\sigma_X$, the
transmittance variance $\sigma_{T}$, and their covariance $Cov\left(T,X\right)$ (equation \ref{E_prop1}-\ref{E_prop4}).

\begin{subequations}
\begin{equation}
\sigma_D^2 =\left|\frac{\partial D}{\partial T}\right|^2\cdot \sigma_{T}^2 + \left|\frac{\partial D}{\partial X}\right|^2\cdot \sigma_X^2 + 2\left|\frac{\partial D}{\partial X}\frac{\partial D}{\partial T}\right|\cdot\text{Cov}\left(T,X\right)\label{E_prop1}\\
\end{equation}
\begin{eqnarray}
\frac{\partial D}{\partial X}&=&-\beta_3\beta_4\frac{T_\infty+T_0}{\left(T_\infty-T+\beta_4X\right)^2}\label{E_prop2}
\end{eqnarray}
\begin{eqnarray}
\frac{\partial D}{\partial T}&=&\beta_3\frac{T_\infty-T_0}{\left(T_\infty-T+\beta_4X\right)^2}\label{E_prop3}
\end{eqnarray}
\begin{eqnarray}
\frac{\partial D}{\partial X}\frac{\partial D}{\partial T}
           &=&-\beta_3^2\beta_4\frac{T_\infty^2-T_0^2}{\left(T_\infty-T+\beta_4X\right)^4}\label{E_prop4}
\end{eqnarray}
\end{subequations}

\subsection{Film handling\label{S_practical}}
\subsubsection{Digitization}
The films were positioned in the middle of an Epson 10000XL flatbed scanner, the 8 inch side is parallel to the long edge of the scanner bed.
Uniformity artifacts due to the position on the scanner are well known\cite{Menegotti_08,Paelinck_07}, therefore, a plastic positioning tool ensures the same position for each scanned film. The films are studied with the orientation mark, clipped by the manufacturer, directed to the lower left corner of the scanner (figure \ref{F_Calscheme}).

To measure the visible light fluence, $\phi$, the scanner is equipped with a transparency cover. Scanning is performed using the Epson scan software, with all corrections turned off. A fixed 8$\times$10 inch area is scanned in 48-bit color mode with a color depth of 16-bit; ie. the pixel values range from 0 to $2^{16}-1$. Files are saved in tagged image file format (TIFF) with 150 dpi resolution. These TIFF images have three layers containing red, green, and blue scan values. All subsequent film processing is performed using home made Matlab scripts \emph{(The MathWorks, Inc)}.

As described in the previous section, transmittance values are determined by the ratio of the fluence transmitted through the film and the fluence incident on the film $T_{\text{RGB}}=\frac{\phi_{t\text{RGB}}}{\phi_{i\text{RGB}}}$. The definition is valid for all three color layers, this is indicated by the RGB index.
$\phi_{i\text{RGB}}$ is measured by a scan without a film the resulting signal is within 1\% of the maximal fluence value ($2^{16}-1$). Therefore, $\phi_{i\text{RGB}}$ is set constant to this value, and :
\begin{equation}
T_{\text{RGB}}=\frac{\phi_{t\text{RGB}}}{2^{16}-1}\label{E_IRGB}
\end{equation}

Our films are scanned at least 48 hours post irradiation, which is the preferred recommendation in the TG 55 report\cite{TG55_1998}. Andr\'es et al.\cite{Andres_10} reported variations in OD smaller than $3\%$ between 2 and 24 hours postirradiation. They report this value for irradiations up to 2.5Gy. Taking in to account that higher doses have a slower density growth\cite{Andres_10}, and the decrease of the subsequent density growth as a function of time, 48 hours between irradiation and read out guaranties differences in density growth below $3\%$.

\subsubsection{Film storage}
The film response is affected by the storage environment, e.g. the impact
of moisture and temperature on optical density\cite{Rink08}. Therefore, all films are separately stored in folders of a file cabinet, which is placed in an air
conditioned environment. The file cabinet limits exposure to
daylight. The separated folders and air conditioning
ensure identical exposure to environmental conditions. The relative moisture and the temperature of the storage room are monitored over the course of a week.

A set of fourteen films was used to assess the effect of the moisture and temperature conditions of the storage environment. The manufacturer supplied the films, stored in vacuum packing. The vacuum pack was chosen to exclude environmental effects during shipping and storage by our local distributer (PEO, Radiation Technology, Hoogstraten,  Belgium). The first seven films (Film$_{V_1-V_7}$) were separated by an interleaving tissue, the last seven were not (Film$_{V_7-V_{14}}$). The vacuum storage was broken on October 20$^{\text{th}}$, 2010. The films were scanned unirradiated on the same day and on subsequent days to evaluate the environmental impact.

For the evaluation of the storage effect, local transmittance variations (variations on short distances -- noise) were not addressed. Therefore, the transmittance layers is sub sampled by taking the average over 5$\times$5 pixel blocks \cite{Zhu_1997}. This operation results in the RGB-transmittance layers denoted by $T_{\text{RGB}}^*\label{E_T*}$. The effect of our local storage conditions is then evaluated by calculating the average of $T_{\text{RGB}}^*$ for the different films, Film$_{V_1-V_{14}}$. The outer 20 pixels were excluded from the calculation to avoid effects of the marks as well as other effects at the film edges. We denote the average and corresponding standard deviation as:
\begin{equation}
\langle T_{\text{RGB}}^*\rangle^*\pm SD\left(T_{\text{RGB}}^*\right)^*\label{E_T*_ave}
\end{equation}
The $\langle.\rangle^*$ and $SD\left(.\right)^*$ notation is used to indicate the difference between the average of a whole film, and the average of a small regions of interest (ROI), denoted by $\langle.\rangle$ and $SD\left(.\right)$ (section \ref{S_DataSets}, equation \ref{E_Tlocal_ave}).

\subsection{A static field calibration methodology}
\subsubsection{The calibration field}
The calibration films are irradiated with 4$\times$4 cm fields with a 6MV photon beam from a Varian Linac
2100C/D. Figure \ref{F_Calscheme} shows a schematic overview of a calibration film. The sizes, the location and orientation of the calibration fields are chosen to maximize the separation between fields. The selected field size (FS) guaranties lateral electronic equilibrium. According to \citeauthor{Todorovic06}\cite{Todorovic06} the calibration curve is not affected by the field size, for field sizes ranging from 2$\times$2 to 10$\times$10 cm.

The protocol uses static fields, to simplify the physics of the calibration fields. For example, by avoiding the use of a
multileaf collimated (MLC) field, leaf transmission effects are excluded.
However, the irradiation of a single calibration field still generates a low dose contribution to the positions of the other fields due to scatter and/or leakage, see section \ref{S_LeakageAndScatter}.

\subsubsection{Transmittance ($\langle T_{\text{RGB}}\rangle$)\label{S_DataSets}}
The calibration film is positioned in a plastic water phantom. The phantom is a stack of twenty 40$\times$40$\times$1cm RW3 plates, which is positioned with 95 cm source to surface distance.  The film was positioned at 100cm from the source. This way there is sufficient
build up and backscatter (5--15 cm).

On the calibration film, eight calibration segments are defined. These eight segments are $1\times1$cm ROI's situated in the center of the $4\times4$cm diamonds as shown in figure \ref{F_Calscheme}. The segments' edges are parallel to those of the diamonds. The segments can be irradiated with a known dose or can be left unirradiated. The average $T$-values of the segments are calculated;
\begin{equation}
\langle T_{\text{RGB}}\rangle\pm SD\left(T_{RGB}\right)\label{E_Tlocal_ave}
\end{equation}

The different calibration segments are irradiated by applying couch shifts. The monitor units (MU) are chosen to have an equal spread between 0 and 100MU, and between 100 and 225MU, see table \ref{T_FilmSumm}. The dose range of the calibration curve is $\sim$ 0.4 to 2.5 Gy, which covers our standard external beam dose range.

\subsubsection{Leakage and scatter conditions ($D_{i\text{ IC}}$)\label{S_LeakageAndScatter}}
In order to account for leakage and scatter contributions to the individual calibration segments, we use the following methodology and assumptions:
\paragraph{}The different segments have symmetric contributions i.e. scatter and leakage contribution of segments 1 to segment 8 is identical to the contribution of segment 8 to segment 1, apart from a scaling factor depending on the delivered dose.
\paragraph{}The ionization chamber (IC) response is weakly dependent on the energy spectrum, therefore, IC-measurements are considered to be accurate enough for this purpose.

For IC measurements, a farmer type IC (FC 65-G TNC, SN 752) is combined with a SI (\emph{Standard Imaging, Inc., Middleton, WI USA}) electrometer (SI SN 070112). The IC is placed in the RW3-phantom, located at the center of the measuring plane, at 100cm from the source.

The eight dose contributions to segment $i$, $D_{i,j\text{ 100 MU}}$, are measured
by placing the IC in the center of segment $i$ while irradiating
100 MU to each segment $j$, $j\in\left[1,8\right]$. Again, couch shifts are used to position the IC in the center of the different segments. The eight measurements should only be performed once, after which only the dose of a reference field should be measured, $D_{i,i\text{Cal 100 MU}}$.

This reference measurement is repeated before the irradiation of a calibration film. Subsequently, the dose delivered to segment i, $D_{i\text{ IC}}$ (equation \ref{E_dosesum2}), is the sum of the contributions of the eight segments on segment $i$, scaled by the ratio of the output measurements ($\frac{D_{i,i\text{Cal 100 MU}}}{D_{i,i\text{ 100 MU}}}$).
\begin{eqnarray}
D_{i\text{ IC}}=\frac{D_{i,i\text{Cal 100 MU}}}{D_{i,i\text{ 100
MU}}}\cdot\sum_{j=1}^{8}{\frac{MU_j}{100MU}\times D_{i,j\text{ 100
MU} }}\label{E_dosesum2}
\end{eqnarray}
Where $MU_j$ denotes the MU delivered to segment $j$.

\subsection{The calibration curve \label{S_calcurve}}
\subsubsection{Location dependence}
Recently \citeauthor{Micke_11} reported a lateral-artefact\cite{Micke_11}, which is a location dependence of the calibration curve of gafchromic films, ie. the $\beta_i$-parameters of the calibration curve are affected by variations of the different dose level positions. To study the impact of such an effect on our calibration methodology, eleven calibration films are irradiated with varying dose levels for the different segment locations. This is done by altering the order of MU used to irradiate the calibration films (Film$_{E_1-E_{11}}$, table \ref{T_FilmSumm}).

Increasing MUs with increasing segment number are used thrice on two different days, Film$_{E_1}$ and Film$_{E_2,E_3}$. Decreasing MUs with increasing segment number is used for Film$_{E_4}$. Film$_{E_5}$ has an arbitrary spread of the MUs on the different locations. Film$_{E_{10}}$ has the same MUs as Film$_{E_4}$ but an arbitrarily segment is left blank and segment six got 225MUs. Film$_{E_6}$ has a low and a high dose field on each row.

\subsubsection{Calibration parameters estimation ($\beta_{i\text{RGB}}$)}
Subsequently, the $\langle T_{\text{RGB}}\rangle-$values, and the corresponding $D_{\text{IC}}-$values, of these eleven films are pooled in ten data sets.

Three pooling methodologies are considered. The first method (data set 1), pools the segments of all eleven films.
The second method (data sets 2-8), pools the segments of a single film: respectively Film$_{E_1}$, Film$_{E_2}$, $\ldots$, Film$_{E_6}$, and Film$_{E_{10}}$.
The third methodology (data sets 9,10), pools the segments of two films, Film$_{E_2}$ + Film$_{E_3}$, and Film$_{E_3}$ + Film$_{E_4}$.

The two film combinations are introduced to increase the number of data points. The Film$_{E_2}$ + Film$_{E_3}$-combination only enlarges the number of data points, while the Film$_{E_3}$ + Film$_{E_4}$-combination ensures both an enlargement of the data set as a varying distribution of low and high dose values on the different segment locations.

For each data set, the $\beta_{i\text{RGB}}$-parameters (equation \ref{E_ItoDose}) are estimated for all three color channels. The estimation is performed in R \emph{(www.r-project.org)}
using a non linear least square fit of equation \ref{E_ItoDose}. The t-statistic and the corresponding two sided p-values are calculated to evaluate the significance of the $\beta_{i\text{RGB}}$-estimations ($\alpha=0.05$).

\subsubsection{Six and Eight segment approach}
Because of film non-uniformities (Figure \ref{F_linescan}) all preceding methods were repeated including only six segments. The segments with $X=-9$cm are avoided in this six segment approach. The methods described before, including all segments, will be referred to as eight segment approach. Only included segments are used in the further evaluation of the approaches (e.g. residual dose error calculation etc.).

\subsection{Error analysis\label{S_error}}
\subsubsection{Residual dose error, $\varepsilon$}
The $\beta_{i\text{RGB}}$-estimations of the previous section are used to convert the $\langle T_{\text{RGB}}\rangle-$values to dose (equation \ref{E_DFilmRGB}). To compare the eight and the six segment approach the residual errors are calculated (equation \ref{E_ErrorFilm}).
\begin{eqnarray}
D_{\text{Film RGB}}&=&\frac{\left(\beta_3\cdot \langle T\rangle-\beta_3\cdot\beta_4 \cdot X \right) - \beta_1}{\beta_2 - \langle T\rangle+\beta_4 \cdot X }\Biggr |_{\text{RGB}}\label{E_DFilmRGB}\\
\varepsilon_{\text{RGB}}&=&D_{\text{Film RGB}}-D_{\text{IC}}\label{E_ErrorFilm}
\end{eqnarray}

The six segment approach on the Film$_{E_3}$ + Film$_{E_4}$ data set has both the lower residuals as the lower p-values for the $\beta_{i\text{RGB}}$-estimations.
Subsequently, the calibration curve resulting from the Film$_{E_3}$ + Film$_{E_4}$ data set is used to calculate the dose values for all the segments of all the films, Film$_{E_1-E_{11}}$. The resulting residual dose errors, $\varepsilon_{\text{RG}}$, are qualitatively compared with an estimation of the dosimetric variation, $\sigma_D$ \ref{E_prop1}. The blue color channel is not evaluated because of the non-significance for the $\beta_{i\text{B}}$-estimations.
\subsubsection{Dosimetric accuracy ($\sigma_{\text{D}}$)}
As described in equation \ref{E_prop1} the dosimetric accuracy consist of components involving positional accuracy ($\sigma_X$), local uniformity or transmittance accuracy ($\sigma_T$), and finally a term taking into account variations of $T$ as function of the $X$-position ($Cov\left(T,X\right)$). No significant variations of the transmittance as function of the $Y$-position were noticed (ANOVA, $\alpha=0.05$). The contributions of the different terms are determined as follows;
\paragraph*{$\sigma_X$:}A 1mm positioning error is assumed both during irradiation as scanning. These two positioning errors are considered to be independent.
\paragraph*{$\sigma_T$:}The local uniformity is defined as the relative standard deviation of $T_{\text{RGB}}$\cite{TG55_1998}, $\frac{SD(T_{\text{RGB}})}{\langle T_{\text{RGB}}\rangle}$. $\sigma_{T\text{ RGB}}$ is estimated by the maximum of the standard deviations of all the segments of all films, Film$_{E_{1}-E_{11}}$:
\begin{equation}
\sigma_{T_{\text{RGB}}}=\max_{\text{Film}_{E_{1}-E_{11}}}\left(SD\left(T_{\text{RGB}}\right)\right)\label{E_sigmaT}
\end{equation}
\paragraph*{$Cov\left(T,X\right):$}Both the relative standard deviation\cite{Zhu_1997} as the variation along the central pixel line\cite{TG55_1998} of $T_{\text{RGB}}^*$ are used to describe the global uniformity:
\begin{equation}
\frac{SD(T^*_{\text{RGB}})^*}{\langle T_{\text{RGB}}^*\rangle^*}
\pm
\frac{\max\left(T^*\right)-\min\left(T^*\right)}{\text{mean}\left(T^*\right)}\biggr|_{\text{central line}}\label{E_centralline}
\end{equation}
The variation of $T$ as function of the $X$-position is estimated by the maximal standard deviation of $T_{\text{RGB}}^*$ of the blank films and the uniformly irradiated film.
\begin{equation}
Cov\left(T_{\text{RGB}},X\right)=\max_{\text{Film}_{E_{7,8,9,11}}}\left(SD\left(T_{\text{RGB}}^*\right)^*\right)\label{E_covTX}
\end{equation}
\section{RESULTS}
\subsection{Transmittance of blankfilms}
\begin{figure}
 \includegraphics[width=7cm]{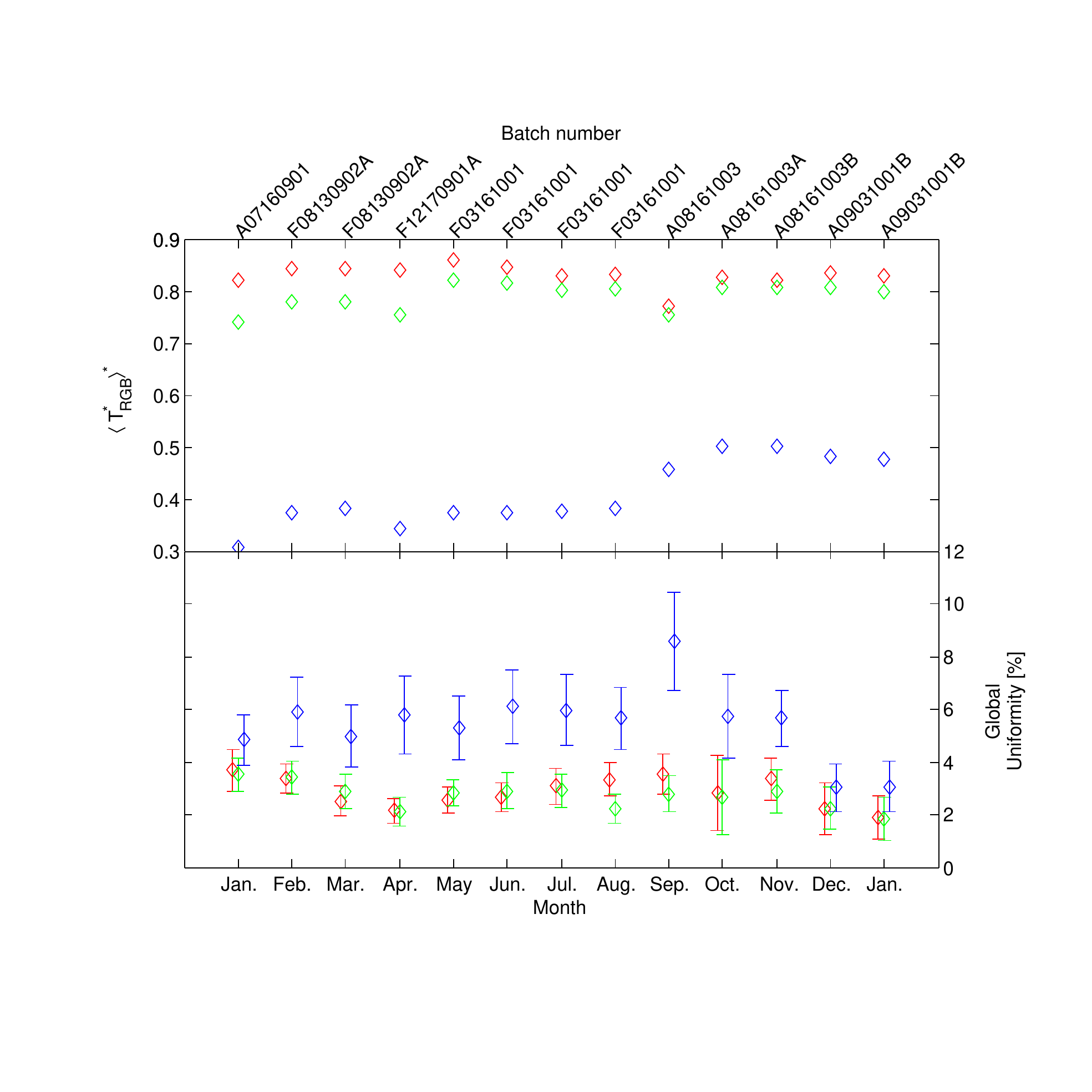}%
 \caption{Blank film stability and uniformity from January 13, 2010 until January 12, 2011. The upper part contains the average transmittance, $\langle T_{\text{RGB}}^*\rangle^*$. \label{F_stability}The global uniformity $\left(\frac{SD\left(T_{RGB}^*\right)^*}{\langle T_{RGB}^*\rangle^*}\pm\frac{\max\left(T^*\right)-\min\left(T^*\right)}{\text{mean}\left(T^*\right)}|_{\text{central line}}\right)$ is illustrated in the lower part.\label{F_stabilityU}}%
\end{figure}

A blank film scan was selected for each month from January 13, 2010 until January 12, 2011. For these films, major changes of both $T^*$ and the global uniformity are noticed, as shown in figure \ref{F_stability}. The red and green channel $\langle T_{\text{RG}}^*\rangle^*$ ranges from 0.77 to 0.86, and from 0.74 to 0.82. The blue channel has a larger range from 0.31 to 0.50. The global uniformity of these film ranges from 2\% to almost 10\%. The last two months an improvement of the blue channel's global uniformity is seen.
\subsection{Association between transmittance and delivered dose}
In a first exploration of the calibration curve, equation \ref{E_ItoDose}, $\beta_4$ is set to zero, see figure \ref{F_AllFit}. All the segments of Film$_{E_{1}-E_{11}}$ are pooled for the $\beta_{i\text{RGB}}$-estimation.
For the red channel all parameters have highly significant results ($p<0.01$); $\beta_1= 1.833\pm0.069$, $\beta_2=0.199\pm0.013$, and $\beta_3=2.159\pm0.083$. The residuals show a slight dependence on the X-location on the film. Additionally a larger spread of the residuals for $X=-9cm$ is noticed.

The green channel has similar results with parameters $\beta_1= 4.487\pm0.340$, $\beta_2=0.084\pm0.040$, and $\beta_3=5.431\pm0.414$ with significance levels $p<2\cdot10^{-16}$, $p<0.05$, and $p<2\cdot10^{-16}$. The green channel residuals have a larger spread for the $X=-9cm$ segments, but the X-dependence as shown in figure \ref{F_AllFit} is not present.

The model was not appropriate for fitting the blue channel data, the $\beta_{i\text{B}}$-estimations are not significant.

\subsection{Storage conditions}
The environmental condition of our storage room is stable with an average relative humidity of 35.0$\pm$4.9\%.
The average temperature is 23.1$\pm$0.2$^{\circ}$C. According to the manufacturer the vacuum packing environment had the following stable conditions; relative humidity = $35-55\%$, and temperature = $18-23^{\circ}$C.

\begin{figure}
 \includegraphics[width=7cm]{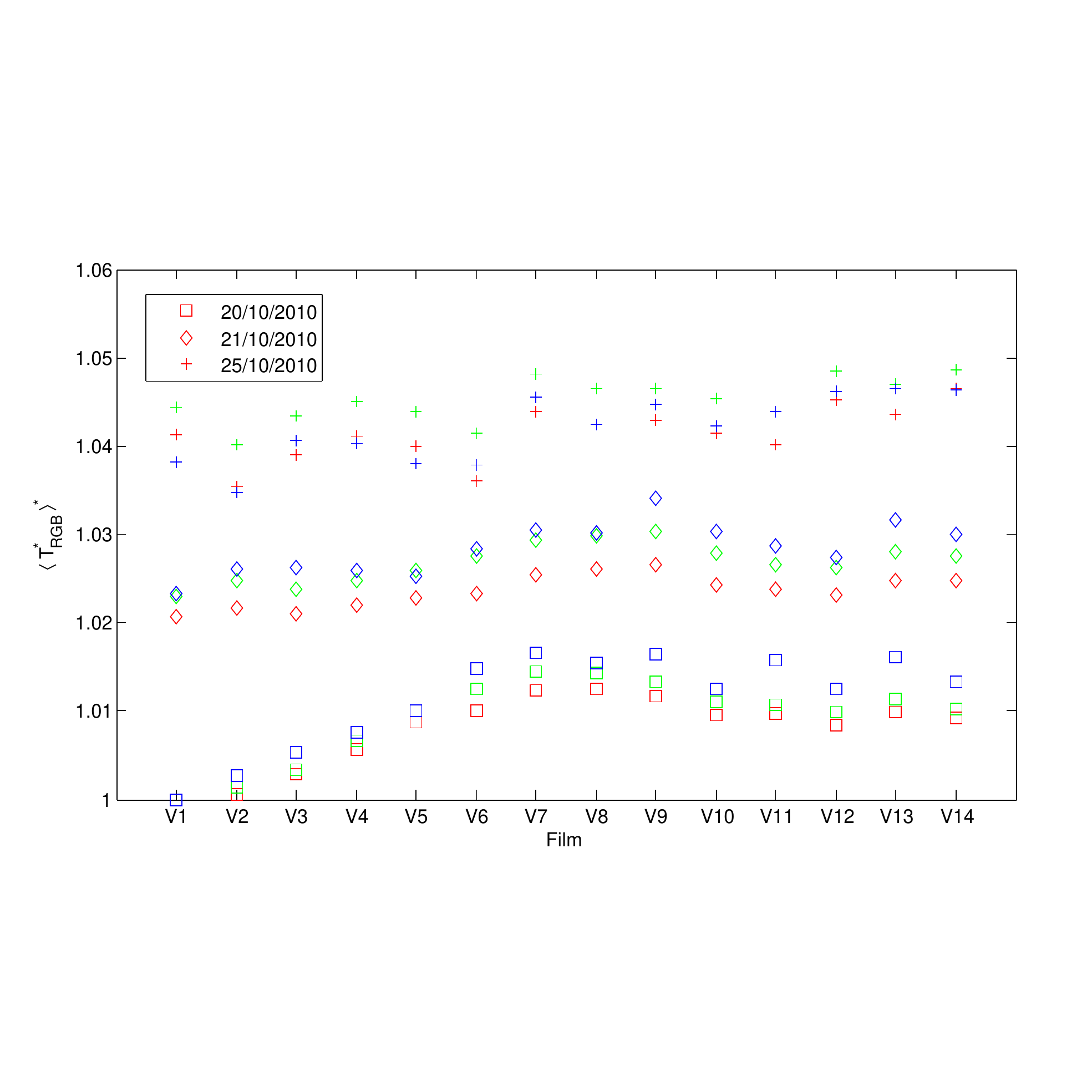}%
 \caption{Effect of environmental conditions of our storage procedure on the vacuum packed films. Average transmittance values, $\langle T_{\text{RGB}}^*\rangle^*$, are presented relatively to the average transmittance of Film$_{V_1}$ on October 20, 2010.\label{F_Vactest}}%
\end{figure}

The evolution of $\langle T^*\rangle^*$ of the vacuum packed films is illustrated in figure \ref{F_Vactest}. There is an increase from Film$_{V_1}$ to a maximal value for Film$_{V_7}$, which is the first film covered with an interleaving tissue. These maximum deviations are $1.25$, $1.45$, and $1.66$\% for the red, green, and blue color channel respectively. The difference is clearly dependent on the location in the stack, and whether an interleaving tissue was used or not.
When the films are stored separately, the impact of the location in the transportation stack and the effect of the interleaving tissue reduces. The differences decrease over time to values smaller than $0.5$\% for the red and green channel and smaller than $1$\% for the blue color channel on the fifth day.
In absolute values,  $\langle T^*_{\text{R}}\rangle^*$ increases from $0.79$ on October 20, 2010 over $0.80$ on October 21, 2010 to $0.81$ on October 25, 2010. The same evolutions are noticed for $\langle T_{\text{GB}}^*\rangle^*$, with values of $0.77$, $0.78$,and $0.80$ (green channel),
and $0.45$, $0.46$, and $0.47$ (blue channel).

\subsection{Six or eight segments\label{S_ResSixOrEight}}
\begin{figure}
 \includegraphics[width=7cm]{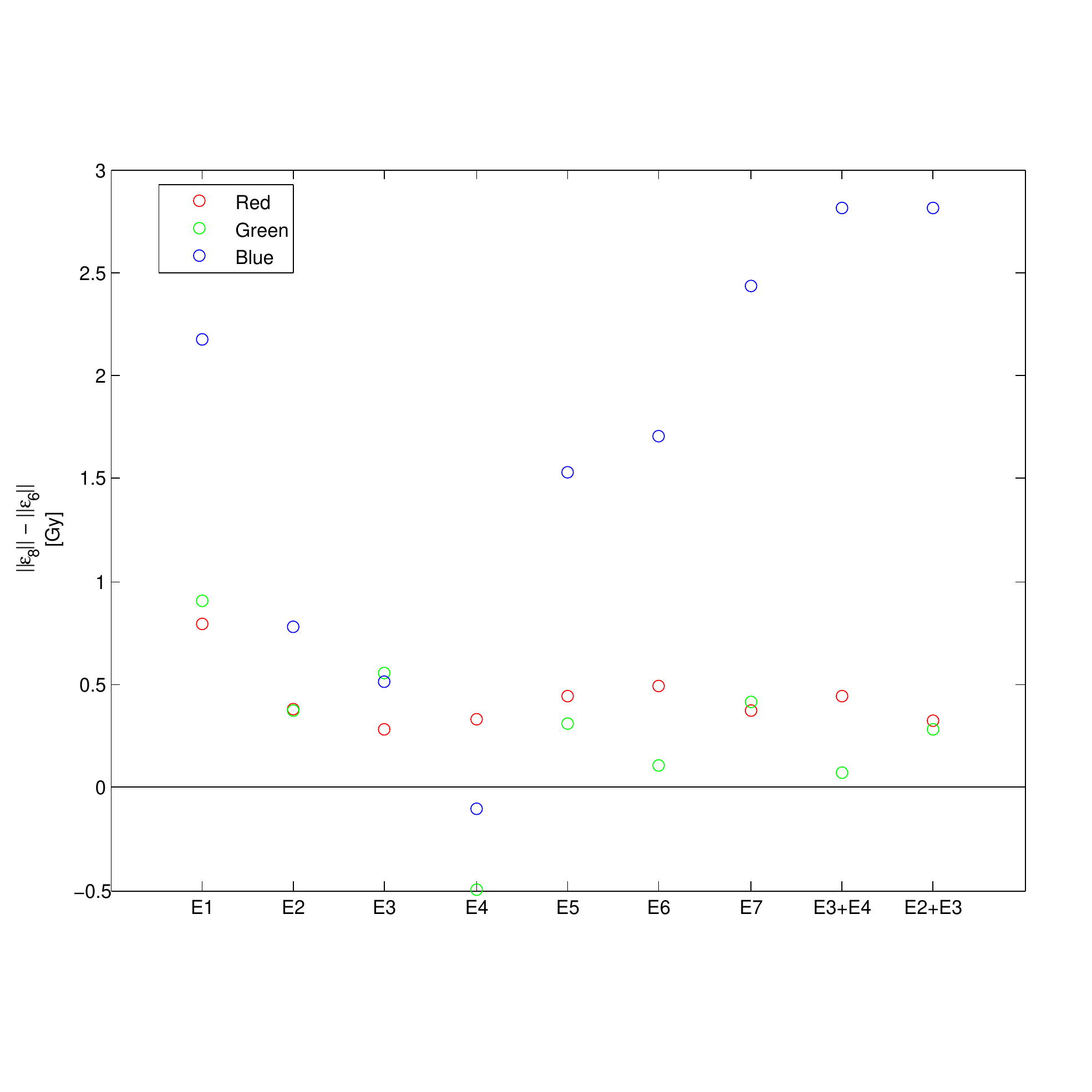}%
 \caption{Eight versus six segment approach. The square root of the squared sum of the residual errors of the six segment approach ($\|\varepsilon_6\|$) is subtracted from that of the eight segment approach ($\|\varepsilon_8\|$).\label{F_ParamSign}}
\end{figure}

The residual errors, $\varepsilon_j$, are calculated for the different data sets, $j$. Only data-sets which are pooled from a single calibration film, or from two calibration films are considered. The difference between the eight and six segment approach is illustrated in figure \ref{F_ParamSign}, using $\|\varepsilon_j\|=\sqrt{\sum\limits_{i\in\text{data set }j}{\varepsilon_{ij}^2}}$.
Avoiding the segments with $x=-9$cm reduces $\|\varepsilon\|$. On average these reductions are $0.43\pm0.15$, $0.28\pm0.38$, and $1.63\pm1.05$Gy for the red, green, and blue channel. An exception for this reduction is found for the film with the opposite MU-order, Film$_{E_4}$. The 33 and 66 MU-segments of Film$_{E_4}$ are located on the $X=-9cm$ position (table \ref{T_FilmSumm}). Eliminating these low dose segments results in an $\|\varepsilon\|$ increase for the green and blue color channel.
\subsection{Calibration}
Only the six segment approach is discussed in this section.
\subsubsection{$\beta_{i\text{RGB}}$-estimations}
Table \ref{T_beta} summarizes the significant $\beta_{i\text{RGB}}$-estimations.
\begin{table*}\caption{Summary of the $\beta_{iRGB}$-estimations.\label{T_beta}}
\begin{center}
\begin{small}
\begin{tabular}{c|c|r@{$\pm$}lc|r@{$\pm$}lc|r@{$\pm$}lc|r@{$\pm$}lc}
  \hline \hline
 \multirow{2}{*}{Film} & & \multicolumn{3}{c|}{$\beta_1$} & \multicolumn{3}{c|}{$\beta_2$} & \multicolumn{3}{c|}{$\beta_3$} & \multicolumn{3}{c}{$\beta_4$} \\
     & & mean&SD & p[\%] & mean&SD & p[\%] & mean&SD & p[\%] & mean&SD & p[\%] \\
    \hline\hline
  \multirow{3}{*}{$E_{5}$} & Red   & 1.64&0.04 & (0.07) & 0.24&0.01 & (0.08) & 1.93&0.06 & (0.08) & (-10&1)$\cdot10^{-4}$ & (0.40)    \\
                           & Green & 4.16&1.43 & (10.0) & 0.13&0.15 & (49.5) & 5.02&1.77 & (10.6) & (- 3&3)$\cdot10^{-4}$ & (42.5)     \\
                           & Blue  & 2.30&0.77 & (9.59) & 0.19&0.05 & (5.90) & 5.64&1.91 & (9.86) & ( -9&1)$\cdot10^{-4}$ & (0.96)     \\\hline
  \multirow{3}{*}{$E_{10}$}& Red   & 2.11&0.09 & (0.18) & 0.15&0.02 & (1.05) & 2.47&0.11 & (0.18) & (-10&1)$\cdot10^{-4}$ & (0.99)     \\
                           & Green & 3.07&0.52 & (2.77) & 0.24&0.06 & (5.85) & 3.67&0.63 & (2.81) & ( -1&3)$\cdot10^{-4}$ & (68.0)     \\
                           & Blue  & 2.11&1.25 & (23.3) & 0.19&0.09 & (15.6) & 5.15&3.05 & (23.4) & ( -6&2)$\cdot10^{-4}$ & (10.5)     \\\hline
  \multirow{3}{*}{$E_{2}+E_3$}&Red & 2.06&0.22 & ($<10^{-3}$) & 0.15&0.05 & (1.21) & 2.41&0.26& ($<10^{-3}$) & (-15&4)$\cdot10^{-4}$ & (0.60)\\
                          & Green  & 3.42&0.82 & (0.31) & 0.19&0.11 & (12.3) & 4.05&0.97 & (0.31) & ( -7&5)$\cdot10^{-4}$ & (22.2)     \\
                          & Blue   & 2.03&1.64 & (25.1) & 0.21&0.12 & (12.5) & 5.01&4.06 & (25.2) & ( -7&4)$\cdot10^{-4}$ & (16.1)     \\\hline
  \multirow{3}{*}{$E_{3}+E_4$}&Red & 1.90&0.13 & ($<10^{-5}$) & 0.19&0.02 & ($<10^{-3}$) & 2.22&0.17& ($<10^{-4}$) & (-11&1)$\cdot10^{-4}$ & ($<10^{-3}$)     \\
                          & Green  & 2.90&0.54 & (0.07) & 0.26&0.06 & (0.28) & 3.43&0.67 & (0.09) & ( -6&2)$\cdot10^{-4}$ & (2.77)     \\
                          & Blue   & 1.43&0.67 & (6.50) & 0.25&0.04 & (0.04) & 3.51&1.68 & (7.08) & ( -8&1)$\cdot10^{-4}$ & (0.06)     \\
      \hline \hline
\end{tabular}
\end{small}
\end{center}
\end{table*}
For the data sets pooled from a single film, significant red channel $\beta_{i\text{R}}$-estimations were found for Film$_{E_5}$ and Film$_{E_{10}}$, the films with an arbitrary mix of low and high dose values on the different locations. Additionally, the blank segment of Film$_{E_{10}}$ resulted in a significant $T_0=\frac{\beta_1}{\beta_3}$-estimation for the green color channel. The green $T_{\infty}=\beta_2$-estimation was borderline significant ($p=0.0585$).

For the calibration sets pooled from two films, the $\beta_{i}$-estimations become highly significant for the red color channel ($p<0.01$). Additionally, a mix of low and high dose values on different locations results in significant estimation of all four $\beta_{i\text{G}}$-parameters (Film$_{E_3}$+Film$_{E_4}$, $p<0.05$).

For this data set, Film$_{E_3}$+Film$_{E_4}$, the blue channel $T_0$ could not be determined, $\beta_{1\text{B}}$ and $\beta_{3\text{B}}$ are not significant, this in contrast to $T_{\infty}=\beta_{2\text{B}}$.
\subsubsection{Calibration curves}
\begin{figure}
 \includegraphics[width=7cm]{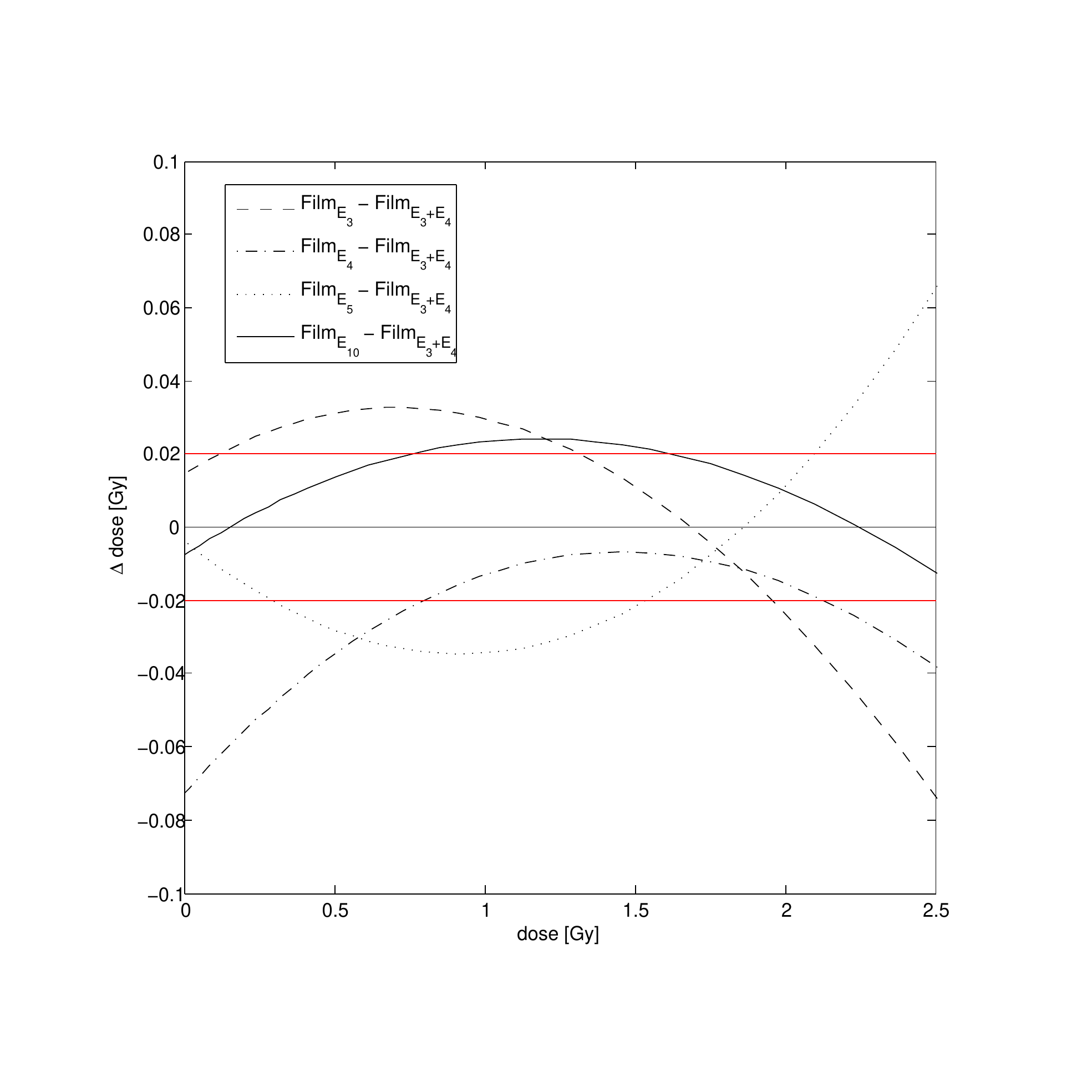}%
 \caption{Comparison of different calibration curves for the red channel transmittance. The curve of the calibration data set pooled from Film$_{E_3}$+Film$_{E_4}$ is considered as the gold standard, this curve is subtracted from the curves from Film$_{E_3-E_5}$, and Film$_{E_{10}}$. The 1\% dose error, compared to 2Gy a common prescribed dose, is illustrated in red.\label{F_Cal_Curve_Compar}}
\end{figure}

In figure \ref{F_Cal_Curve_Compar} the calibration curves resulting from the data sets pooled from Film$_{E_3-E_5, and E_{10}}$ are compared with the calibration curve of Film$_{E_3}$+Film$_{E_4}$. The results are evaluated with a 0.02Gy threshold, 1\% of 2Gy, a commonly prescribed dose. Large differences ($>$0.02Gy) between the curves are noticed. Film$_{E_{10}}$ has the best performance. Film$_{E_3}$ (increasing MU, with increasing segment number) under estimates the dose for values larger than 2Gy and over estimates the dose between 0.1 and 1.4Gy. Film$_{E_4}$ (opposite MU-order of film Film$_{E_3}$) has an overall underestimation of the dose ($\Delta<$0Gy), the low dose range (values $<$0.5Gy) has the worst performance. Film$_{E_5}$ (arbitrary MU-order) is comparable with Film$_{E_3}$ but over estimations are seen where Film$_{E_3}$ has underestimations and visa versa. Film$_{E_{10}}$ (=Film$_{E_4}$, but with a blank segment and higher dose for segment 6) has the better performance for the whole dose range.
\subsection{Error analysis}
\subsubsection{Transmittance accuracy}
\begin{figure}
 \includegraphics[width=7cm]{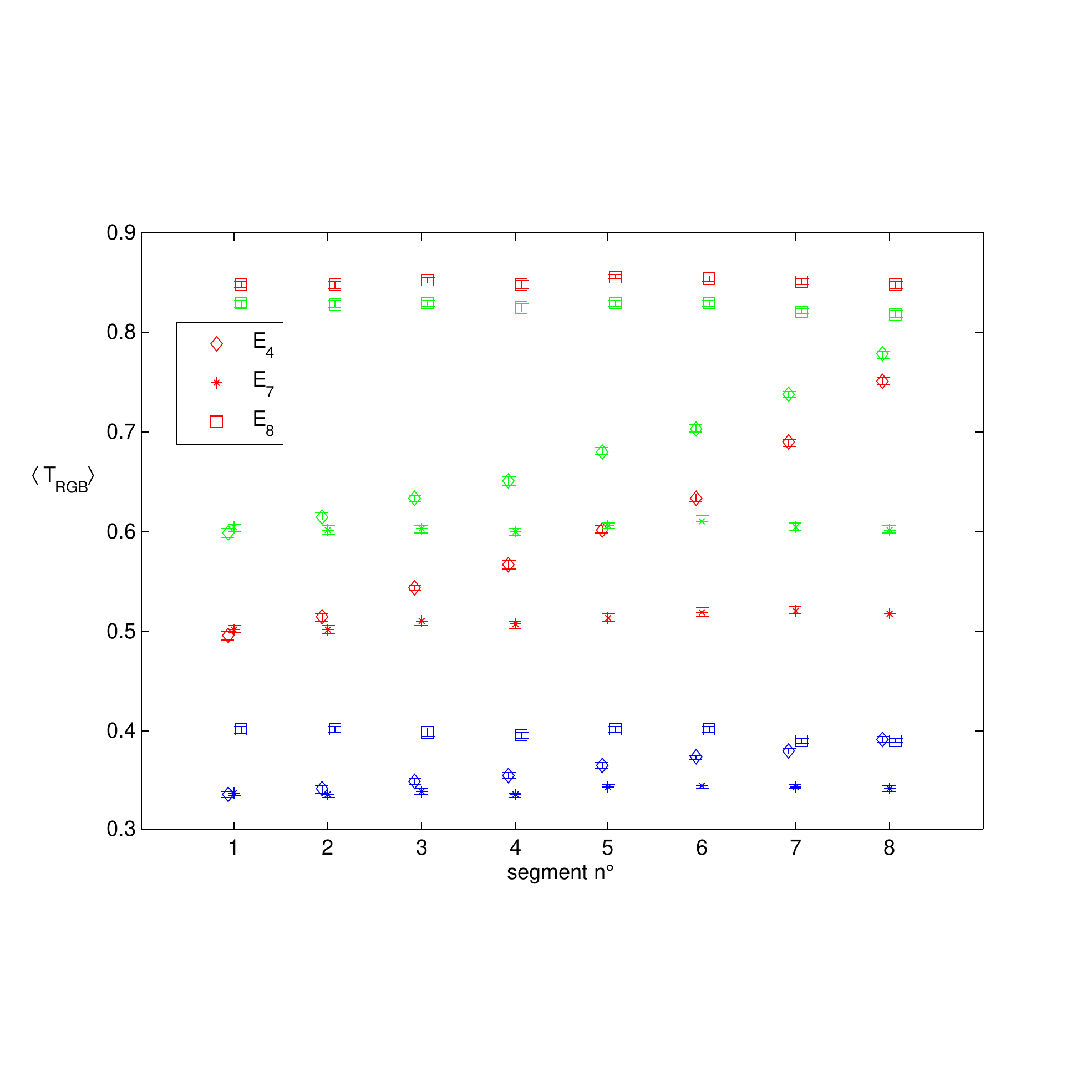}%
 \caption{Local uniformity: the segments' average transmittance and corresponding standard deviation ($\langle T_{\text{RGB}}\rangle \pm SD\left(T_{\text{RGB}}\right)$ ) are shown for the following films: a calibration film (Film$_{E_4}$), a uniformly irradiated film (Film$_{E_7}$), and
 a blank film (Film$_{E_8}$).\label{F_ULocal}}%
\end{figure}

\begin{figure}
 \includegraphics[width=7cm]{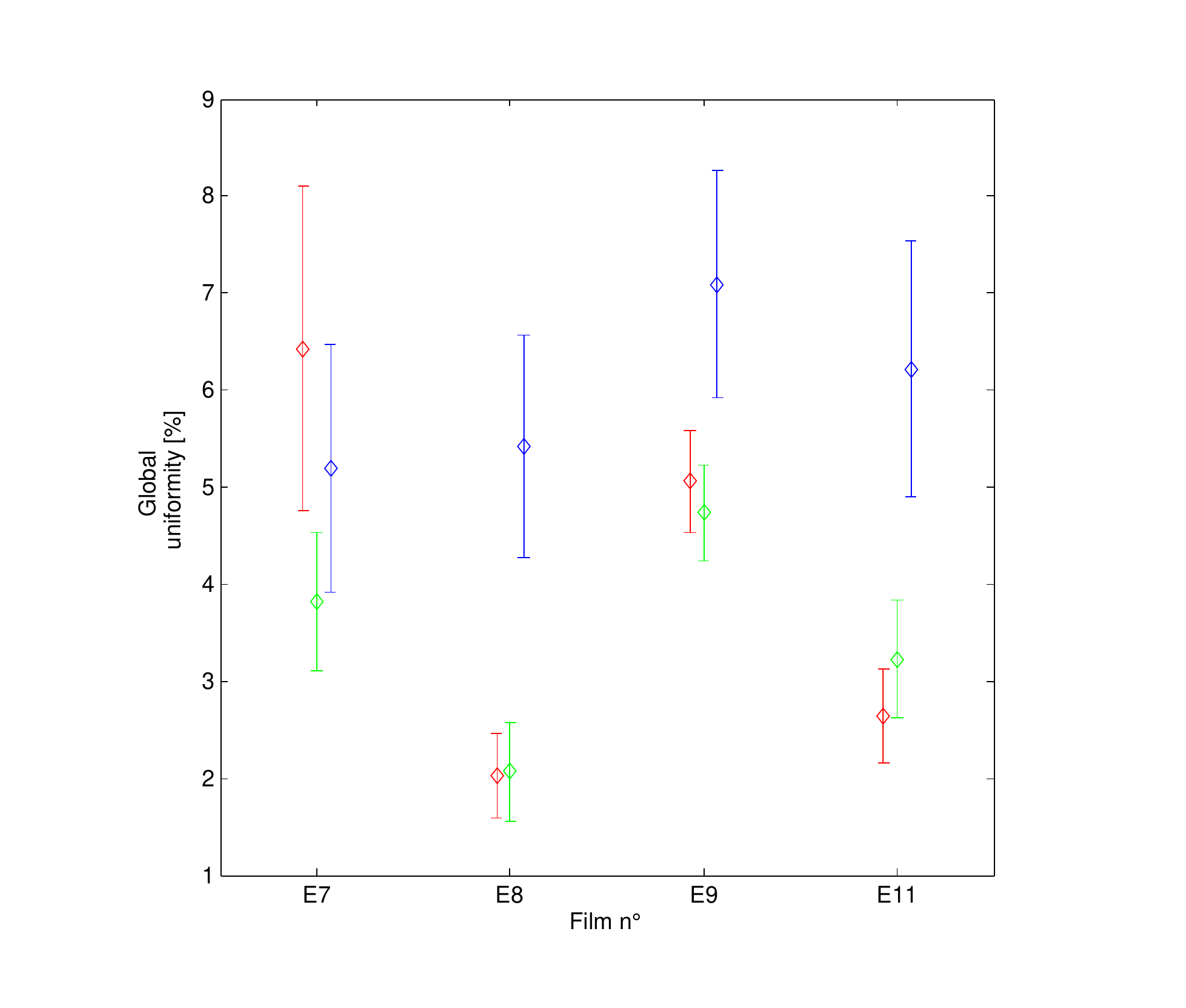}%
 \caption{Global uniformity $\left(\frac{SD\left(T_{RGB}^*\right)^*}{\langle T_{RGB}^*\rangle^*}\pm\frac{\max\left(T^*\right)-\min\left(T^*\right)}{\text{mean}\left(T^*\right)}|_{\text{central line}}\right)$ of a uniformly irradiated film (Film$_{E_7}$) and three blank films (Film$_{E_{8,9,11}}$).\label{F_UGlobal}}%
\end{figure}

Figures \ref{F_ULocal} and \ref{F_UGlobal} show respectively the local and global uniformity. The local uniformity is smaller than $1.08\%$, $0.97\%$, and $1.15\%$, for the red, green and blue channel. The global uniformity ranges from 2$\%$ to $7\%$. The blue channel has worse global uniformity than the red and green channel.
The transmittance covariations with the $X$-position ($\text{Cov}\left(T,X\right)$) are $5.7\cdot10^{-3}$,  $5.1\cdot10^{-3}$, and $5.5\cdot10^{-3}$ (red,green,and blue channel) for the six segment approach. Compared to the eight segment approach, the six segment approach has  a 33\% reduction of the red channel $\text{Cov}\left(T,X\right)$. The green and blue channel covariances of the different approaches are comparable. $\sigma_T$ was not found to be different for the two approaches.
\subsubsection{Residual dose error and dosimetric accuracy ($\varepsilon$ and $\sigma_{\text{D}}$)}
\begin{figure}
 \includegraphics[width=7cm]{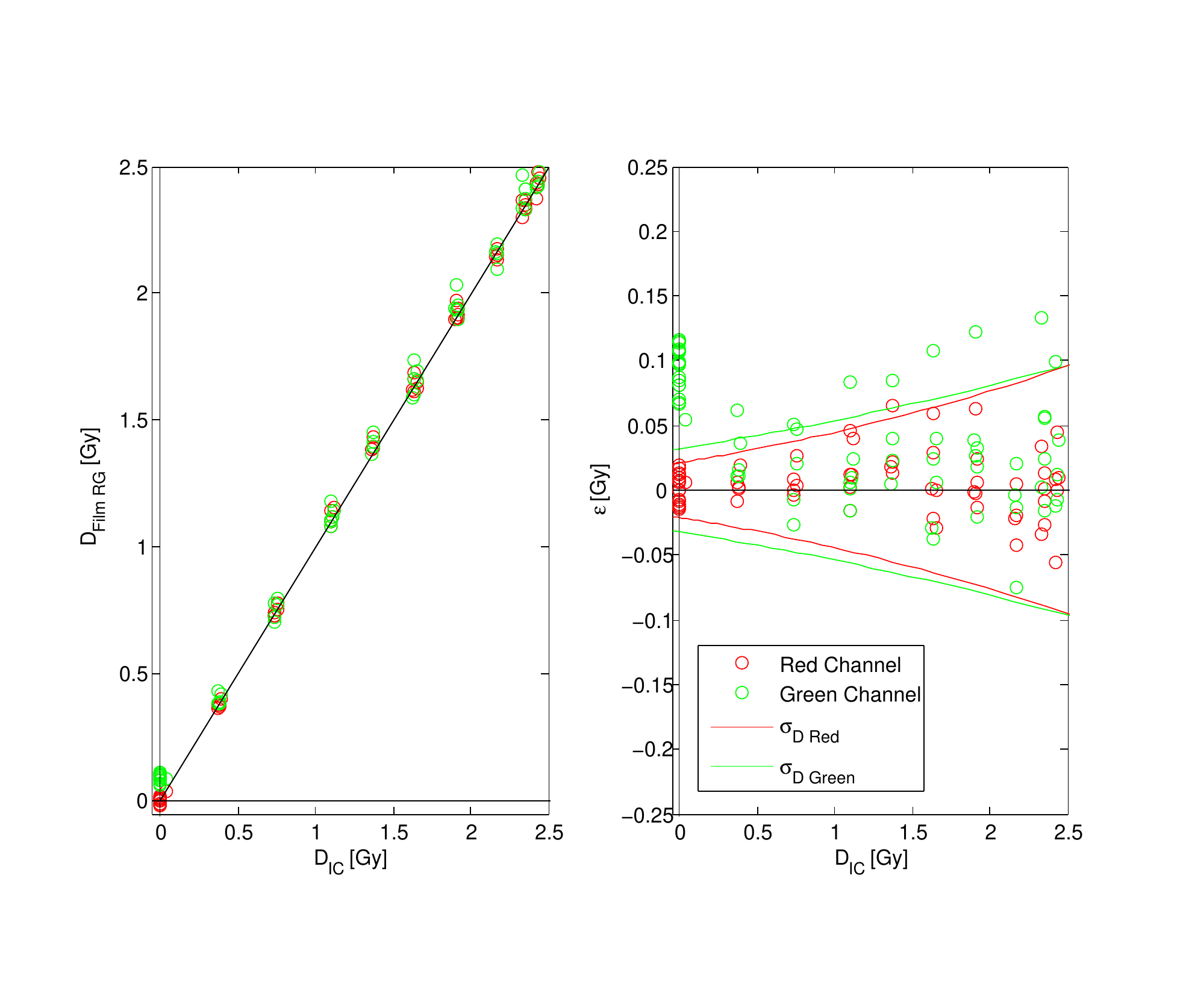}%
 \caption{Left, calculated dose values of Film$_{E_1-E_{11}}$ using calibration curve based on $E_3+E4$ (six segment approach). Right, dose differences ($\varepsilon$) of the left figure and dose standard deviation ($\sigma_D$) calculated using equations \ref{E_prop1} till \ref{E_prop4}.\label{F_measure8}}%
\end{figure}

Figure \ref{F_measure8} illustrates the dose values based on the calibration curve of the Film$_{E_3}$+Film$_{E_4}$-data set (six segment approach).
The dose from the film measurements, $D_{\text{Film RGB}}$, and the corresponding IC measurements, $D_{\text{IC}}$, are shown in the left figure. On the right side, the residual errors are displayed. An over estimation is found in the green channel for the un-irradiated segments.
In the $[0.04,2.5]$Gy dose range, excluding all unirradiated segments, the red channel residual errors deviate 0.62$\pm$1.79\% (mean$\pm$1SD) from the expected dose, $D_{\text{IC}}$, with a range of $[-2.3\%,4.9\%]$. In the same dose range the green channel residual errors deviate 2.1$\pm$3.63\% from the expected dose. These residual errors have a $[-3.6\%,6.9\%]$ range, with three outliers of 7.6, 9.53, and 16.6\%.

The dosimetric accuracy, $\sigma_D$, is calculated for $X=0$cm. For the red channel the calculated dosimetric accuracy is a good estimate for the upper limit of the residual errors. For the green channel the calculated $\sigma_D$ is an under estimation of this upper limit.
\section{DISCUSSION}
\subsection{$T_0$}
\subsubsection{Stability}
Figure \ref{F_stability} illustrates an instability for the gafchromic EBT 2-films used in our department. The $T_0^*$-values are largely dependent on the production (variation between different batches). The instability is present in all three color channels but it has different effects, it stresses the necessity to calibrate each batch. An advantage of our calibration approach is the fact that the estimated calibration parameters indicate the zero dose transmittance ($T_0$), and therefore, inform the users about the changes of the film.

For the same films, global non-uniformities were noticed. To work with a classic optical density protocol, a location dependent $\phi_i^*$-estimation should be made for each bach (f.e. Garc\'ia-Gardu\~no\cite{Garcia_10} et al. proposes an average scan of five randomly selected blank films as $\phi_i^*$-estimation for gafchromic EBT films). With a throughput of more than one box of films each month our protocol needs to be as transparent as possible. Therefore, we do not prefer the cumbersome $\phi_i^*$-estimation, and $T$ is used instead of OD.
\subsubsection{Storage}
A stack of films without interleaving tissues allows the films to slide over each other, and static electric charges are created on the film. The resulting electrostatic forces stick the films to each other. This way the films are only exposed to the environment at the film edges. The interleaving tissue, avoids the direct contact of the films and serves as a guide for the humidity (David Lewis, ISP\cite{ISP}, private communication). This is shown in figure \ref{F_Vactest}, where the films separated by interleaving tissues have similar $T_0$-values, and the films without interleaving tissue have increasing $T_0$-values in function of the location in the stack. When exposed to our local environmental conditions the films tend to reach a stable state, relative to each other, after 1 day. Meanwhile, the manufacturer reintroduced the interleaving tissue for the transport of the films.
\subsection{Calibration curve}
\subsubsection{Protocol}
Only a correlation between $\varepsilon$ and the $X$-location was noticed (figure \ref{F_AllFit}). The absence of a correlation between $\varepsilon$ and $D_\text{IC}$ validates the assumption of using only eight IC measurements to describe the low dose contributions.

The low local uniformity values (figure \ref{F_ULocal}) support the choice of the (small) $4\times4$ calibration fields. On the other hand, because of the absence of a $\sigma_T$-difference between the six and the eight segment approach, $\sigma_T$ is not affected by the global non-uniformities. Therefore, the $1\times1$ cm segment are small enough to evaluate local uniformity.

To cope with global non uniformities a location dependent extension is added to the rational function, additionally the six segment approach was introduced. This approach decreases the degrees of freedom (six vs eight data points per film for four $\beta_{i\text{ RGB}}$-parameters). Nevertheless, this approach is an improvement for the protocol, indicated by the reduction of $\|\varepsilon\|$, figure \ref{F_ParamSign}.

Figure \ref{F_Cal_Curve_Compar} illustrates the better performance of the calibration curve from Film$_{E_{10}}$. The improved significance of $\beta_{i\text{G}}$-parameters of Film$_{E_{10}}$ compared to Film$_{E_5}$ indicate an improvement of the protocol due to the introduction of a no dose segment. The necessity of low dose values is further supported by the Film$_{E_4}$-results, where the absence of low dose data points in the six segment approach enlarges the residual error, $\|\varepsilon\|$. This puts the dose levels used in the protocol in question. Future improvements of our protocol are the introduction of 0 MU segment, and the other MU values will be selected to have equal intervals between the $\langle T_{\text{R}}^*\rangle^*$ data points.

\subsubsection{Location dependence}
Remarkable are the significance levels for the $\beta_{i \text{RGB}}$-parameters. For the single film calibrations (Film$_{E_1-E_6,E_{10}}$), only the films with an arbitrary spread of the MUs on different locations generate significant results for the red channel.

For the two film methodology, the Film$_{E_3}$+ Film$_{E_4}$-data set, results in significant $\beta_i$-estimations for \emph{both} the red and the green color channel. This in contrast to the Film$_{E_2}$+ Film$_{E_3}$-data set, where only the red $\beta_{i}$-estimations are significant. The difference between these two data sets is the MU-order used to irradiate the films. Film$_{E_3}$+ Film$_{E_4}$ has an opposite MU-order on the two films while Film$_{E_2}$+ Film$_{E_3}$ has the same MU-order on the two films. Therefore, the spread of high and low dose segments on different positions is found to be relevant for the calibration of the gafchromic EBT 2-films.

In other words, the calibration curve is location dependent ($D=f\left(T,X,Y\right)$), which is also mentioned by \citeauthor{Micke_11}\cite{Micke_11}. The study of such an effect, combining the dose and the X-location (e.g. a $\beta_5 X\cdot D$-term), was outside the scope of is this work. However, using two films with a spread of different dose levels on different locations, makes our calibration curve an acceptable surrogate for such a location dependent calibration curve.

\subsection{Error analysis}
\subsubsection{Transmittance accuracy}
Although, $T$ is used instead of $OD$, it is interesting to compare the uniformity results with literature values. After all, both $T$ and $OD$ form the input of a calibration curve, and the error on these input will determine the error of the resulting dose values.

Excellent local uniformity values were found ($<1.15\%$) which is lower than $3\%$ and $5\%$ relative standard deviations for the MD-55-2-film listed in the TG 55 report \cite{TG55_1998}. Zhu et al. \cite{Zhu_1997} reports $7\%$-$15\%$ global OD-variations for the MD-55 film, which is comparable with our transmittance results ($\pm7\%$). Nevertheless, adaptations were necessary to cope with the global non uniformity (six segment approach, and $X$-dependent parameter in the calibration curve).

\subsubsection{Residual dose error and dosimetric accuracy ($\varepsilon$ and $\sigma_{\text{D}}$)}
In a valid calibration protocol $\varepsilon$ should have a normal distribution with a zero mean and a standard deviation from the previous section ($\varepsilon_D \sim \mathbf{N}\left(0,\sigma_D\right)$). This relation is not evaluated because of the small amount of data, $n=6\sim8$ per dose level. Therefore, we have chosen an over estimation of $\sigma_D$, introduced by using the \emph{maximum} of the local and global uniformity variations in equations \ref{E_sigmaT} and \ref{E_covTX}.

The red color channel $\sigma_D$-prediction was able to estimate the maximal dose errors, figure \ref{F_measure8}.

For the green channel the calculated $\sigma_D$-values are not sufficient to estimate the maximal dose errors in the protocol. Neither was the protocol capable to convert $T_{\text{G } 0}$ to dose, see figure \ref{F_measure8}.
To cope with this problem, blank segments could be introduced in the calibration protocol. The significance of the $\beta_{1\text{G}}$ and $\beta_{3\text{G}}$-estimations for the film with a blank segment (Film$_{E_{10}}$) supports this assumption.
\section{CONCLUSION}
The proposed calibration protocol for absolute gafchromic film dosimetry has a straightforward association between transmittance and dose, equation \ref{E_ItoDose2}. The approachable protocol required only eight preparation measurements, and a single reference measurements on the day of calibration. The use of simple static 4$\times$4 fields is suited for this purpose even as the use of IC-measurements to map the low dose contributions.

Film changes are reported both, due to the production process as due to the storage environment. The strength of our protocol is that each calibration characterizes the physical parameters of the films, $T_0$, $T_\infty$, and a factor scaling the impact of the dose ($\frac{1}{\beta_3}$).

However, because of non-uniformities the original intended association between $T$ and $D$ required an adaptation. An additional term ($\beta_4\cdot X$) and the avoidance of certain segments, are positively evaluated as improvements of the protocol.

The protocol requires a spread of low and high dose segments on two calibration films and the segments with a higher spread of the residual dose errors should be excluded. Additionally the use of a blank segments is advisable.

All color channels are dose dependent, but a trustworthy calibration protocol with error prediction was only found for the red channel in the $]0,2.5]$Gy dose range. In this dose range the red channel dose error range equals $[-2.3\%,4.9\%]$. For the red color channel an upper limit for the dose error could be predicted using an error propagation analysis. The green color channel has the higher residual errors, and requires more attention to the estimation of $T_0$. The blue color channel is dose dependent but could not be calibrated with the proposed protocol.\\

\subsubsection*{Summary of our current calibration protocol}
\paragraph*{Preparation measurements}Eight preparation measurements are performed once to estimate the low dose contributions.
\paragraph*{Storage}The films are stored in separate folders of a file cabinet to ensure the same environmental exposure off all films.
\paragraph*{Blank Check}After a few days of storage, for stabilization purposes, all films are scanned unirradiated. Thereafter, an automated script calculates  $\langle T_{RGB}^*\rangle^*\pm SD\left(T_{RGB}^*\right)^*$, as well as the local uniformity off all the films.
\paragraph*{Dose levels}A blank segment is introduced for a better estimation of $T_{0}$. In our current protocol the dose range is extended from 0 to $\sim$ $4.5$Gy , using MU ranging from $0$ to $393$. Where the number of MUs are chosen to be equally distributed in the red transmittance domain.
\paragraph*{2 Films}Two calibration films are irradiated using opposite MU-orders.
\paragraph*{Calibration Check}Before using clinical films, the calibration films  are scanned and evaluated.
\paragraph*{Segment Selection}Based on the Blank Check and the Calibration Check a decision is made wether certain segments should be avoided or not (e.g. segments with $X=-9$cm). In our latest clinical evaluations such a segment avoidance was not necessary (batch Nr. A09031001B, A11011001, A12171002B).
\paragraph*{48h post irradiation}We wait 48 hours before scanning the films.
\paragraph*{Scan all films in one session}
All the films that need to be evaluated are scanned successively, including the calibration films.
\paragraph*{Convert to dose}
A calibration curve is created for the scan session and all the films are converted to dose.
\paragraph*{Calibrate each box of films}
\paragraph*{\textbf{Acknowledgements}} Mr. K. Poels is acknowledged for extensive discussions. The authors like to thank ISP and PEO for their fruitful discussion and their support in this work.
\appendix
\section{Transmittance vs dose\label{S_A1}}
Equation \ref{E_ItoDose} (\ref{E_A1}) can be transformed to \ref{E_DosetoI} (\ref{E_A2}) using the following routine juggling.
\begin{small}
\begin{eqnarray}
T &=&\frac{ \beta_1 + \beta_2 \cdot D }{\beta_3 + D}+\beta_4\cdot X\label{E_A1}\\
\left(T-\beta_4\cdot X\right)\cdot \left(\beta_3 + D\right) &=&\beta_1 + \beta_2 \cdot D \nonumber\\
\left(T-\beta_4\cdot X\right)\cdot \beta_3 + \left(T-\beta_4\cdot X\right)\cdot D &=& \beta_1 + \beta_2 \cdot D \nonumber\\
\left(T-\beta_2-\beta_4\cdot X\right)\cdot D &=& \beta_1 - \beta_3\cdot\left(T-\beta_4\cdot X\right) \nonumber\\
D &=& \frac{\beta_1 - \beta_3\cdot T +\beta_3\beta_4\cdot X}{T-\beta_2-\beta_4\cdot X} \nonumber\\
D &=& \frac{\beta_3\cdot T -\beta_3\beta_4\cdot X -\beta_1}{\beta_2-T+\beta_4\cdot X} \label{E_A2}
\end{eqnarray}
\end{small}
These equations can be reduced further using three physical parameters. The first two parameters characterizes the film using the zero and infinite dose transmissions ($T_0$ and $T_{\infty}$). The third parameter, $\beta_3$ scales the impact of the dose.
\begin{small}
\begin{eqnarray}
&\left\{
\begin{split}
T &=\frac{ \beta_1 + \beta_2 \cdot D }{\beta_3 + D}+\beta_4\cdot X\nonumber\\
D &= \frac{\beta_3\cdot T -\beta_3\beta_4\cdot X -\beta_1}{\beta_2-T+\beta_4\cdot X}\nonumber\\
\end{split}
\right.&\\
&\LARGE{\Downarrow}&\nonumber\\
&\left\{
\begin{split}
T &=\frac{ \frac{\beta_1}{\beta_3} + \beta_2 \cdot \frac{D}{\beta_3} }{1 + \frac{D}{\beta_3}}+\beta_4\cdot X\nonumber\\
D &= \beta_3\cdot\frac{T -\beta_4\cdot X -\frac{\beta_1}{\beta_3}}{\beta_2-T+\beta_4\cdot X}\nonumber\\
\end{split}
\right.&\\
&\LARGE{\Downarrow}&\small \left\{\begin{split} T_0&=\frac{\beta_1}{\beta_3},\text{ for }\beta_4=0&\\
T_\infty&=\lim_{D\rightarrow\infty}\left(\frac{\beta_1 + \beta_2 \cdot D}{\beta_3 + D}\right)&\\
   &=\beta_2\text{\hspace{1eM}(l'H\^opital's rule)}
\end{split}\right. \nonumber\\
&\left\{
\begin{split}
T &=\frac{ T_0 -T_\infty+T_\infty+ T_\infty \cdot \frac{D}{\beta_3} }{1 + \frac{D}{\beta_3}}+\beta_4\cdot X\nonumber\\
D &=\beta_3\cdot\frac{T -\beta_4\cdot X -T_0}{T_\infty-T+\beta_4\cdot X}\nonumber\\
\end{split}
\right.&\\
&\LARGE{\Downarrow}&\nonumber\\
&\left\{
\begin{split}
T &=\frac{ T_0 -T_\infty+T_\infty\cdot\left(1+ \frac{D}{\beta_3}\right) }{1 + \frac{D}{\beta_3}}+\beta_4\cdot X\nonumber\\
D &=\beta_3\cdot\frac{T -\beta_4\cdot X -T_0}{T_\infty-T+\beta_4\cdot X}\nonumber\\
\end{split}
\right.&\\
&\LARGE{\Downarrow}&\nonumber\\
&\left\{
\begin{split}
T&=T_\infty + \frac{T_0 - T_\infty }{1+\frac{D}{\beta_3}}+\beta_4\cdot X\label{E_StartForDelta}\\
D&=\beta_3\cdot\frac{\left(T_0+\beta_4\cdot X\right)-T}{T-\left(T_\infty+\beta_4\cdot X\right)}\nonumber
\end{split}
\right.&\\
&\LARGE{\Downarrow}&\beta_4=0\nonumber\\
&\left\{
\begin{split}
T&=T_\infty + \frac{T_0 - T_\infty }{1+\frac{D}{\beta_3}}\label{E_StartForDelta}\\
D&=\beta_3\cdot\frac{T_0-T}{T-T_\infty}\nonumber
\end{split}
\right.&\\
\end{eqnarray}
\end{small}

\end{document}